\documentclass{aipproc}

\layoutstyle{6x9}

\usepackage{bm}

\title[Time variation of fundamental constants]%
	{Time variation of
	fundamental constants: two phenomenological models
	\footnote{Dedicated to Augusto García  and Arnulfo Zepeda,
	fellow scientists and friends}}

\author{H\'ector Vucetich\footnote{On leave of absence,
	Observatorio Astron\'{o}mico, Universidad Nacional de La
	Plata, Paseo del Bosque S$/$N, CP 1900 La Plata,
	Argentina.}}{address={Instituto de F\'\i{}sica, UNAM},
	email={vucetich@fisica.unam.mx}}

\newcommand{\logder}[1]{\ensuremath{\dot{#1}/{#1}}}
\newcommand{\hlogder}[1]{\ensuremath{\logder{#1}/H_0}}
\newcommand{\spader}[1]{\ensuremath{c^2/g\mid\nabla{#1}\mid/{#1}}}

\begin{document}

\begin{abstract}
	Time variation of fundamental constants, a strong signal of
	new physics, is analyzed comparing extant data with two
	phenomenological models. Results are discussed.
\end{abstract}

\maketitle

\section{Introduction}
\label{Intro}

     The Standard Model of Fundamental Interactions (SM) together with
General Relativity (GR) provides a consistent description of all known
low energy phenomena (i.e.  low compared with the Grand Unified (GU)
energy scale) in good agreement with experiment. These theories depend
on a set of parameters called the ``fundamental constants'', which are
supposed to be universal parameters; i.e.: time, position and
reference frame invariant.  Indeed, Einstein Equivalence Principle, on
which GR is based, implies such an invariance and thus the detection
of a time, position or reference system dependence of any of the
fundamental constants imply the existence of ``new physics'', beyond
SM or GR.

     Thus, variation of fundamental constants has been an active
subject of research since the introduction of the Large Number
Hypothesis (LNH) by Dirac \cite{Dirac37,Dirac38} long ago. (See also
\cite{BT86}). Several of these theoretical approaches are briefly
discussed in the next section.

     Partially inspired by these theoretical results, many attempts
have been made to detect through observation or experiment a variation
of fundamental constants, since this latter will produce a host of
different phenomena. Reference \cite{TVOFC1} discusses critically
several of the above results and modern reviews can be found in
references \cite{Landau:2000cc,Uzan:2002vq}. Table \ref{sample} shows
a summary the most accurate results obtained from several sources,
assuming that the fine-structure constant is the only one which varies
in time.

\begin{table}
\begin{tabular}{lccr}
Phenomenon & Age ($10^9$ y) &  $\logder{\alpha}$ (y$^{-1}$) &
	Reference\\ 
\hline
Rb vs. Cs Clocks & 0 & $(4.2 \pm 6.9) 10^{-15}$ &
	\cite{Sortais01} \\
Oklo reactor & 1.8 & $(-0.2 \pm 0.8) 10^{-17}$ &
	\cite{Fujii:2002hc,Fujii:1998kn} \\
Age of Earth & 4.5 & $(1.0 \pm 0.8) 10^{-15}$ & \cite{TVOFC1} \\
FS in QAS & 12.3   & $(-0.4 \pm 1.1) 10^{-15}$ &
	\cite{2001MNRAS.327.1237M} \\
MM in QAS & 12.4 & $(-0.61 \pm 0.23) 10^{-15}$ & \cite{Webb:2000mn} \\
HFS H vs Opt in QAS & 11 & $(0.32 \pm 0.50) 10^{-15}$ &
	\cite{1995ApJ...453..596C} \\
CMB Fluctuations & 15	& $(0.0 \pm 1.3)  10^{-12}$ &
	\cite{LandauRec01,Martins02} \\
\hline
\end{tabular}
\caption{Sample results on the time variation  of  the fine structure
constant $\alpha$. The columns show the phenomenon used, the time from
the present of the observation, the estimated rate of change and the
reference.}\label{sample} 
\end{table}

	The recent announcement of a detection of a nonzero variation
of the fine structure constant $\alpha$ in the absorption spectra of
distant quasars \cite{Webb:1998cq,Webb:2000mn,2001MNRAS.327.1208M} has
started an active experimental and theoretical research activity
around this subject and we shall try to discuss a few of the current
issues.

	The organization of this paper is as follows: next section
briefly discusses theoretical foundations of the problem and then the
effective low energy model used for testing purposes is
described. Modern tests are briefly reviewed and finally results and
conclusions are stated.

\section{A zoo of competing theories}
\label{Zoo}

	Variation of fundamental constants is a common prediction of a
large set of theories which can be divided in two main groups:
\begin{enumerate}
\item Those attempting to unify gravitation with all the other
fundamental interactions through the introduction of additional
dimensions (Superstrings, Kaluza-Klein).
\item Those implementing some version of Mach's Principle.
\end{enumerate}

	In the following, I shall try to extract the common features
of these theories leading to the prediction of time or space variation
of fundamental constants.

\begin{description}
\item[Kaluza-Klein theories] These theories, the simplest attempting a
unification of gravitation and other fundamental interactions
\cite{Kaluza21,Klein26}, assume that there are $D$ additional
compactified (internal) dimensions with a scale $R \sim \ell_P$.  The
equations of motion for gravitation and matter can be deduced from
Einstein's equations in $4 + D$ dimensions:
\begin{equation}
R_{MN}^{(D)} - \frac{1}{2} g_{MN}^{(D)} R^{(D)} = \frac{8\pi G_D}{c^2}
	T_{MN}^{(D)}  
	\label{equ:Eins-D}
\end{equation}
where the sub(super)script $D$ indicates the $4 + D$ dimensional
tensors. 

	The fundamental constants appear, in these theories, as
functionals of the additional dimensions geometry when the system is
reduced to a 4-dimensional one \cite{Chodos80,Wein83,Marciano84}. One
expects, in general, for any of the SM gauge-fields coupling constants:
\begin{equation}
\alpha_i = C\left(\frac{R_0}{R}\right)^{-2}
		\label{equ:KK-alpha}
\end{equation}
where $R_0$ is the present value of the internal space. Since
generic cosmological solutions of equations \eqref{equ:Eins-D} have 
time-varying $R(t)$, Kaluza-Klein theories predict time-variation of
fundamental constants. Indeed, $R$ can be parametrized in the form $R
= \ell_P e^\phi$, where $\phi$ is called the dilaton field and is a
general function of space and time.

\item[Superstrings]
	Superstring theories \cite{GSW-St} assume that there are extended
fundamental objects, called strings, that move in $4 + D$ space and
whose excitations represent (rather, would represent) the basic fields
and particles. It is assumed that the low energy form of the theory is
similar to a Kaluza-Klein theory. The existence of a dilaton entails a
variation of fundamental constants \cite{Damour:1994ya,Damour:1994zq}.

\item[Machian theories]
	The purpose of this rather large class of theories is to
implement Mach's principle \cite{Mach95}: 
\begin{em}
	The mass $M$ of a physical body is a functional $M[x\mid\rho]$
	of the mass distribution $\rho(x)$ in the Universe.
\end{em}
The same ideas has been extended to other conserved quantities such as
charge or even parameters like light speed.
\begin{description}
\item[Variable mass theories] In these theories
\cite{BD61,HN66,Bekenstein:1977rb,Wesson84,Ma90a,WesKill1,WesKill2},
the mass of a particle is proportional to a field $m = m_0 \mu(x)$,
whose source is the mass density in the Universe $\rho(x)$.
\item[Variable charge theories] Particles charges are assumed to be
functions of a universal field $\epsilon(x)$
\cite{Beckens82,Bekenstein:2002wz,SBM02,OP02} whose source is related
to the charge distribution in the Universe $e = e_0
\epsilon(x)$. Maxwell's equations take their usual form, but with a
modified field tensor:
\begin{equation}
F^{\mu\nu} = \frac{1}{\epsilon}
	\left[\partial^\mu\left(\epsilon A^\nu\right) - 
	\partial^\nu\left(\epsilon A^\mu\right) \right]
					\label{F-Modif}
\end{equation}
	It can be shown that gauge invariance is preserved and charge
	conservation holds.
\item[Varying speed of light theories] Here light speed $c = c_0
\psi(x)$ is a variable field
\cite{Moffat93a,Moffat93b,Albrecht99,Barrow99,Magueijo00a,Magueijo00b},
and so space and time derivatives do not commute. Maxwell's equations
take the form:
\begin{equation}
\frac{1}{c} \partial_\mu \left(cF^{\mu\nu}\right) = 4 \pi j^\nu
					\label{Mod-Maxwell}
\end{equation}
that implies charge non-conservation \cite{Landau:2001gz}.
\end{description}
\end{description}

\section{Effective theories}
\label{Eff}

	Both Superstring and Kaluza-Klein theories reduce, in the low
energy limit, to effective theories of the Machian
type. Schematically, the reduction is as follows:
\begin{equation}
\mbox{Superstring} \Rightarrow \mbox{Kaluza-Klein} \Rightarrow
	\mbox{Machian} \label{Reduc}
\end{equation}

	The latter theory is defined through the effective Lagrangian
	density
\begin{equation}
\mathcal{L}_\mathrm{eff} = \sqrt{-g} \left[\mathcal{L}_G +
	\frac{\omega}{2} \left(\partial_\mu \phi\right) +
	f(\phi)\mathcal{L}_M\right] 
					\label{L_eff}
\end{equation}
where $\mathcal{L}_G, \mathcal{L}_M$ are the gravitational and matter
lagrangians, $\phi$ is the ``Machian'' field (dilaton) and $f(\phi)$
represents a (set of) coupling function(s). This lagrangian is assumed
valid for energies below the GUT scale. The effective gauge
coupling constants will have the form:
\begin{equation}
\alpha(x) = \frac{f[\phi(x)]}{f[\phi(x_0)]} \alpha(x_0)
					\label{alpha_eff}
\end{equation}
where $x_0$ is a reference point. Besides, the laboratory scale
constants are related to the GUT scale constant through the
renormalization group equations \cite{Wein83,Marciano84}.

	  The cosmological equations for the above effective theory
have the form (for the present Universe):
\begin{eqnarray}
\frac{1}{H_0^2}\left(\frac{\dot{R}}{R}\right)^2 = 
	\frac{\Omega_0}{R^3} + \Omega_\Lambda +
	\Omega(\phi,\dot{\phi}) \nonumber\\
\ddot{\phi} + 3 \frac{\dot{R}}{R} \dot{\phi} = - f'(\phi) T(R,\phi)
					\label{Eff-Cosmo}
\end{eqnarray}
and thus time variation of the fundamental constants is a generic
prediction. Depending on the details of the theory, oscillations in
the fundamental constants are possible \cite{Morikawa,Osc}.

	Local equations of motion for the effective theory are
obtained linearizing them around the reference point $x_0$: $\phi =
\phi_0 + \varphi$. One obtains for a quasistatic field:
\begin{equation}
\nabla^2 \varphi = 4 \pi G_N \lambda_D \rho 		\label{Eff-Local}
\end{equation}
using the fact that the source of the $\varphi$ field is
related to the trace of the energy momentum tensor. The parameter
$\lambda_D$ is related to the local value of $f'$. The local Machian
field is then proportional to the newtonian potential
$U_G(\bm{r})$ and the theory predicts space variation of
fundamental constants:
\begin{equation}
\alpha_i = \alpha_i^0 \left[1 + \mu_i \frac{U_G(\bm{r})}{c^2} \right]
							\label{Eff-Sp-Var}
\end{equation}
which in turn implies position-dependent binding energies.

\section{Observational testing}
\label{Test}

	The variation of fundamental constants produces a host of
observable effects that can be used to test or search for the effect. 

\begin{description}
\item[Short time $(t \ll H_0^{-1})$ effects]
	Among the local effects that can be used to test effective
theories, change in planetary parameters (such as radii and moments of
inertia) and orbital evolution are important, albeit low accuracy
tests \cite{TVOFC1,TVOFC2,Osc}. Laboratory experiments based on the
comparison of different clock rates are presently more accurate
\cite{Turneaure83,Godone:1993zw,Prestage95,Sortais01}. The Oklo
phenomenon (a natural reactor that operated about 2 Gy ago) provides
the most accurate bounds on the local time variation of fundamental
constants \cite{Damour:1996zw,Fujii:1998kn,Fujii:2002hc}. Finally, the
coincidence of different determinations of the age of the Earth
($\alpha$ vs $\beta$ ages) puts useful limits on the local changes of
the constants \cite{TVOFC1}.

\item[Spatial effects]
	The classical Dicke argument \cite{Dicke59,Haughan79,Uzan:2002vq}
shows that a spatial variation of the fundamental constants would
entail the appearance of an anomalous gravitational acceleration
detectable in E\"otv\"os-like experiments. This can be easily seen
from \eqref{Eff-Sp-Var} since energy-dependent binding energies
imply that the energy of a macroscopic body in a gravitational field
will have a passive gravitational mass different from its inertial mass
which in turn implies an anomalous acceleration in a gravitational
field:
\begin{equation}
\bm{a} = \left( 1 + \sum_i \mu_i \frac{E_{Bi}}{c^2} \right)
\bm{g}
					\label{Anom-Accel}
\end{equation}
which can be tested in E\"otv\"os-like experiments.

\item[Cosmological effects]
	There have been several attempts to directly measure the time
variation of fundamental constants in the absorption spectra of
distant quasars, which originate in cold gaseous clouds interposed
between the quasar and ourselves. There are measurements of fine
structure
\cite{1995ApJ...453..596C,1996AstL...22....6V,2001MNRAS.327.1237M},
yielding directly $\Delta \alpha / \alpha$. The use of multiple
transitions from several elements in the same clouds has improved the
sensibility of the method, yielding the first positive detection of a
change in $\alpha$
\cite{Webb:1998cq,Webb:2000mn,2001MNRAS.327.1208M}. 

Other quantities
measured are $x = \alpha^2 g_p \frac{m_e}{m_p}$ from the ratio of
optical transitions to the 21 cm radiation
\cite{Wolfe76,1979AJ.....84..699W,1995ApJ...453..596C}; $y = g_p
\alpha$ from rotational transition frequencies in diatomic molecules
to the 21 cm radiation \cite{Murphy01c} and $\mu = \frac{m_e}{m_p}$
from molecular hydrogen absorption \cite{1998ApJ...505..523P}. This
latter result is the only direct result on the Higgs sector of the
SM. 

%
%
\begin{table}
\begin{tabular}{lcccr}
\hline
\tablehead{5}{c}{c}{From local data}\\
\hline
Model & \hlogder{\alpha} & $\hlogder{G_F}$ & $\hlogder{\Lambda_{\rm Q}}
	$ & Ref.\\ 
\hline
Phenomenological & $2\cdot10^{-4} $ & $ 2\cdot10^{-1} $ & $
1.7\cdot10^{-2} $ & \cite{TVOFC1}\\
\hline
\tablehead{5}{c}{c}{From cosmological data}\\
\hline
Model & \hlogder{\alpha} & $\hlogder{G_F}$ & $\hlogder{\Lambda_{\rm Q}}
	$ & Ref.\\ 
\hline
Kaluza-Klein & $7\cdot10^{-8}$ & $3\cdot10^{-7}$ & $3\cdot10^{-6}$&
	\cite{Landau:2000cc} \\
Bekenstein & $3.2\cdot10^{-8}$ & & &  \cite{Landau:2000cc}\\
\hline
\tablehead{5}{c}{c}{Spatial bounds}\\
\hline
Model & \spader{\alpha} & \spader{G_F} &
	\spader{\Lambda_{\rm Q}} & Ref.\\
\hline
Phenomenological & $1.4\cdot10^{-8}$ & $3.3\cdot10^{-2}$ &
	$1.0\cdot10^{-8} $ & \cite{Chamoun02}\\
\hline
\tablehead{5}{c}{c}{Conservation Laws}\\
\hline
Model & \hlogder{c} & &	 & Ref.\\ 
\hline
VSL & $4\cdot10^{-16}$ & & & \cite{Landau:2001gz}\\
\hline
\end{tabular}
\caption{Bounds on time and space variation of fundamental
constants. Temporal bounds are in units of the Hubble constant $H_0$
while spatial bounds are in units of the local Newtonian acceleration
$g/c^2$.}
\label{tab:Result}
\end{table}

\item[Early Universe Results] Both the Cosmic Microwave Backgroud
(CMB) spectrum \cite{Battye:2000ds,Avelino:2001nr,LandauRec01} and
primordial ncleosynthesis \cite{BBFHe,Landau:2000cc} have been used to
constrain time variation of fundamental constants, since well measured
results are affected by their changes.

\item[Conservation laws]
	The structural changes implied by the trasnformation of
constant parameters into scalar fields can destroy the validity of
well known conservation laws. Such is the case of the conservation of
electric charge, that is violated in the VSL theories proposed in
\cite{Albrecht99,Barrow99}. Indeed, from \eqref{Mod-Maxwell} the
following law for the variation of charge in a closed system can be
derived \cite{Landau:2001gz}
$	
\frac{\dot{Q}}{Q} = - \frac{\dot{c}}{c}		\label{Q-NonCons}
$	
\end{description}

\section{Results}
\label{Concl}

\begin{figure}
\includegraphics{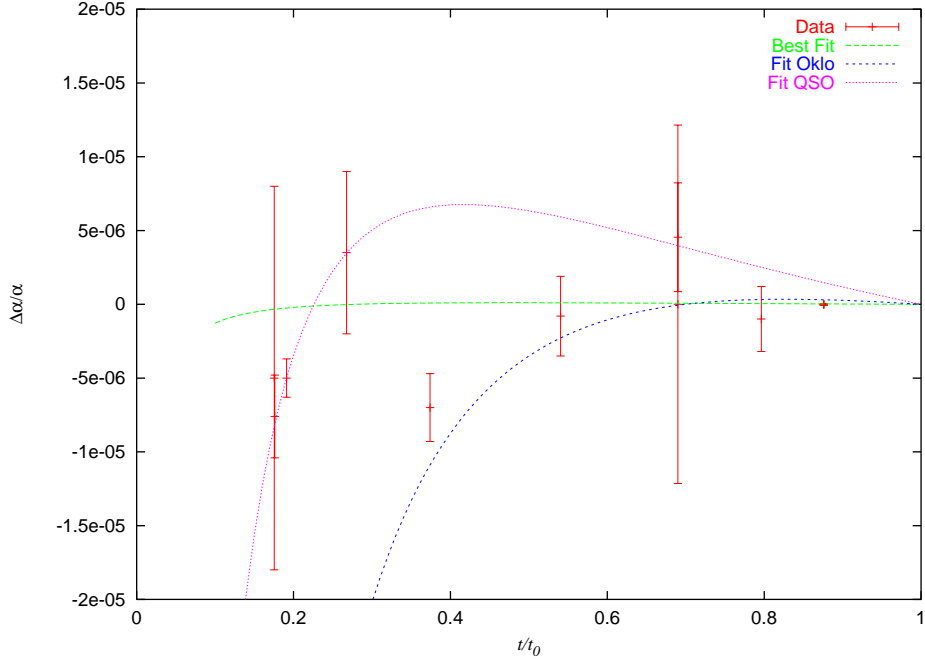}
\caption{Comparison of observational cosmological data with
Bekenstein's model. Data from references \cite{Fujii:1998kn} and
\cite{2001MNRAS.327.1208M} cannot be reconciled. The ``Best Fit'' is a
least squares one, with $\sqrt{\omega}/l_P = 0.15$; the other two have
been forced to $\sqrt{\omega} = l_P$.}
\label{fig:Compare}
\end{figure}

	Although limits on the time variation of fundamental constants
can be obtained from individual measurements, as in table
\ref{sample}, an analysis of the full wealth of data is necessary in
order to compare theory and experiment \cite{TVOFC1,Landau:2000cc}.

	In reference \cite{Landau:2000cc} such an analysis was carried
out for Kaluza-Klein theories and the Bekenstein theory. This
complements the model independent analysis of references
\cite{TVOFC1,TVOFC2}. Table \ref{tab:Result} shows the upper limits
found for the time and space variation of fundamental constants in
those and similar analysis. 

	Our results are consistent with no time variation of
fundamental constants over cosmological times, as most of
observational results indicate. Indeed, excluding the positive data
points of references
\cite{Webb:1998cq,Webb:2000mn,2001MNRAS.327.1208M}, no significant
change of the final results is found. Besides, the full set of
E\"otv\"os-like experiments are consistent with null results
\cite{Chamoun02} and so is electric charge conservation with the
constancy of $c$ \cite{Landau:2001gz}.

\section{Conclusions}

	 The full data set analyzed shows null results for the time
variation of fundamental constants in the nearby time. On the other
hand, neither Kaluza-Klein theories nor Bekenstein-like ones can
explain the full set of cosmological observations. The Bekenstein
model is specially interesting since it is similar to the low energy
limit of the dilatonic sector of superstring theories.  Figure
\ref{fig:Compare} shows a few attempts to explain the data of
references \cite{Webb:1998cq,Webb:2000mn,2001MNRAS.327.1208M} and
\cite{Damour:1996zw,Fujii:1998kn,Fujii:2002hc} with a Bekenstein
model: no combination of parameters can do it. The difficulties are
even worser if the spatial limits are included in the data set (See,
however, reference \cite{Bekenstein:2002wz}).

	We conclude that none of the available theories is able to
explain the full data set on time and space variation of fundamental
constants, specially the quasar results. Were these proved free of
systematic errors, we should conclude that, together with the
existence and nature of dark mass and energy, they herald a
forthcoming crisis on our current theoretical ideas.

\section{Acknowledgements}

The author acknowledges the Organizing Board for their gentle
invitation, Instituto de F\'i{}sica, UNAM for support, Observatorio
Astron\'omico, University of La Plata for permission, J. Bekenstein,
J. D. Barrow and D. F. Torres for useful comments on the manuscript
and Chaac, the Mayan god of rain, for  keeping weather fine.

\bibliography{gener,cosmo,tvref_th,tvref_ph}

\end{document}